\documentclass[aps,floats,showpacs,epsf,twocolumn]{revtex4}
\usepackage{amsfonts}
\usepackage{graphics}
\usepackage{graphicx}
\begin{document}

\title{Magnetic field induced inequivalent vortex zero modes in strained graphene}

\author{Bitan Roy$^{1,2}$}

\affiliation{$^1$ Department of Physics, Simon Fraser University, Burnaby, British Columbia, Canada V5A 1S6 \\
$^2$ National High Magnetic Field Laboratory, Florida State
University, FL 32306, USA}

\date{\today}

\begin{abstract}

Zero energy states in the Dirac spectrum with U(1) symmetric massive vortices of various underlying insulating orders in strained graphene are constructed in the presence of the magnetic field. An easy plane vortex of antiferromagnet and quantum spin Hall orders host two zero energy states, however, with two different length scales. Such \emph{inequivalent} zero modes can lead to oscillatory charge and magnetization, and their usual quantizations get restored only far from the vortex core. Otherwise, these zero modes can be delocalized from each other by tuning the mutual strength of two fields. One can, therefore, effectively bind a single zero mode in the vortex core. A possible experimental set up to capture signature of this theory in real graphene as well as in optical honeycomb lattices is mentioned. Generalization of this scenario with underlying topological defects of Kekul\'{e} superconductors can localize a single Majorana mode in the vicinity of the defect-core.    
\end{abstract}


\maketitle

\vspace{10pt}

\section{Introduction}

The existence of states at precise zero energy in the spectrum of a relativistically invariant Dirac equation with underlying topological defects \cite{Jackiw-topo}, recently attracted ample interest, following the successful fabrication of \emph{graphene}. The extent of the spectrum over the positive and the negative energies provides robustness to such states at the middle of the spectrum against \emph{weak} local perturbations. The number of zero energy states, however, depends only on the universal properties: total magnetic flux enclosed by the system \cite{aharanov-Jackiw-zeroenergy}, kink of the mass \cite{semenoff-domainwall}, and the vorticity of a mass vortex \cite{Jackiw-topo}. For example, a half-filled Landau level at zero energy, leads to the formation of Hall plateaus at fillings $\nu=\pm 2$ \cite{graphene_QHE}. The existence of zero energy modes can lead to fractionalization of quantum numbers, e.g., \emph{charge} \cite{chamon-hou-mudry}, Majorana modes in the core of the superconducting vortex \cite{ghaemi} or half-vortex \cite{herbut-roy-half-vortex} and additional competing order parameters in the core of the vortex \cite{herbut-topo, ghaemiPRB}. The Kosterlitz-Thouless (KT) scaling of longitudinal resistivity ($R_{xx}$) in neutral graphene strongly suggests the possibility of vortex excitations in graphene subject to the magnetic fields \cite{ong}. Vortex-zero modes in the Dirac spectrum can exist in the presence of either real \emph{or} pseudo magnetic fields \cite{herbutmassmagnet}. The nature of the electronic ground state in the presence of fictitious gauge field lately gained attention following the experimental realization of the strain induced fields \cite{Levy, aswin, roy-thesis}. In the presence of uniform real ($B$) \emph{and} pseudo ($b$) magnetic fields, quantized Hall conductivity in graphene is expected to discern plateaus at \emph{all} integer values of $e^2/h$, however, at \emph{noninteger fillings}, when the former is kept stronger. The plateau of Hall conductivity with $\sigma_{xy}= e^2/h$, can be formed by developing a \emph{ferrimagnet} order, due to distinct degeneracies of two inter-penetrating relativistic Landau levels living in the vicinity of two inequivalent Dirac points. Otherwise, their degeneracies are proportional to the effective gauge fields near the Dirac points, $B \pm b$, respectively \cite{bitan-oddQHE}. \\

A fundamental question is, therefore, raised: Does the vortex like defect have any zero energy mode when graphene is subject to both the real \emph{and} the pseudo magnetic fields ? Deposition of graphene on a metallic substrate at relatively high temperatures, followed by cooling, produces a strain induced pseudo (axial) magnetic field due to the mismatch in compressibility with the substrate \cite{Levy}. This system can then be placed in a real magnetic field. Recently, a hexagonal honeycomb lattice has been realized in an optical lattice system (OLS)\cite{optical-lattice}. One can, therefore, study the proposed scenario in honeycomb OLS as well. In the OLS, a magnetic field is introduced via synthetic gauge potentials, and its strength can be varied over a wide range \cite{Y.-J.Lin}. On the other hand, a pseudo magnetic field can arise from specific modulations of the nearest-neighbor hopping (NNH) amplitudes  \emph{only} \cite{herbut-pseudo, guinea, aswin, roy-thesis}. \\

In graphene possible candidates with requisite U(1) symmetry are antiferromagnet (AF), Qquantum spin Hall orders projected onto the easy plane and the Kekul\'{e} bond density wave (KBDW). The former two can acquire an easy plane in the magnetic field due to the Zeeman coupling \cite{herbut-s03}. Otherwise, AF and QSH phases are energetically favored by strong onsite \cite{herbut-juricic-roy} and second-neighbor repulsions, respectively \cite{raghu}. KBDW can be stabilized by electron-phonon interactions \cite{chamon-magneticfield}. However, a QSH phase can be realized even at sufficiently weak next-neighbor repulsion when a fictitious magnetic field penetrates graphene \cite{herbut-pseudo,roy-thesis}. Immaterial of the vortex nature of Dirac quasi-particles we find that there are always \emph{two} zero energy modes. However, categorically two different situations arise: When the underlying order parameters do not couple two inequivalent Dirac points, e.g., AF and QSH orders, zero energy modes have two different characteristic lengths. With underlying KBDW order, two zero energy modes differ from the ones in the absence of pseudo fields only by an overall factor, hence, enjoy a unique length scale. In the former situation, we show that the zero modes can be delocalized from each other arbitrarily by tuning the relative strength of the fields \emph{only}. The present mechanism can, therefore, allow one to localize a single zero mode near, while pushing the other one far from, the vortex-core. The charge and the magnetization then exhibit \emph{oscillatory} behavior, while restoring their usual quantization only at large distances. A previous study shows that in the absence of gauge fields the zero energy modes have the same decay lengths, hence, exact quantization is restored everywhere \cite{herbut-AFvortex}. Additionally, if there exists a half-skyrmion of AF order, however weak, which otherwise splits the zero modes if uniform, maximally lowers the energy by filling the state with shorter length scale. The zero energy states can be computed exactly if the fields are uniform, although they always exist for any arbitrary flux profile and/or shape of the vortex mass. Tuning the strength of the synthetic gauge fields and/or modulated NNH, thereby of pseudo magnetic fields, one can grasp peculiarities of these zero modes on honeycomb OLSs.   \\

The rest of paper is organized as follows. In the next section, we write down the free Dirac Hamiltonian subject to both real and pseudo magnetic fields and mention the single-particle Landau-level spectrum. Existence of zero energy modes in the presence of an underlying vortex with easy-plane AF order is discussed in Sec. III. Robustness of the zero modes against any modulation of the field and/or the vortex mass profile is presented in Sec. IV. Section V is devoted to study the internal structure of the two-dimensional zero energy manifold and various competing orders in the core of the vortex. Mid-gap states with underlying vortices of easy-plane QSH order and KBDW order are respectively considered in Secs. VI and VII. Concluding remarks and a discussion on related issues are presented in Sec. VIII.  \\

\section{Free Hamiltonian with real and pseudo magnetic field}

To describe the massless, chiral Dirac fermionic excitations around the two inequivalent corners of the Brillouin zone, suitably chosen at $\vec{K}_1 =-\vec{K}_2= (1,1/\sqrt{3}) (2\pi/a \sqrt{3})$ \cite{semenoff}, we construct an eight-component spinor $\Psi(\vec{x})=\left(\Psi_\uparrow(\vec{x}), \Psi_\downarrow(\vec{x}) \right)^\top$, with $\Psi^\top_\sigma (\vec{x})=\left[ (u^\dagger _{1\sigma} (\vec{x}), v^\dagger_{1\sigma} (\vec{x}), u^\dagger _{2\sigma} (\vec{x}), v^\dagger _{2\sigma}(\vec{x}) )\right]$. $\sigma=\uparrow,\downarrow$ stands for electrons spin projection along the $z-$direction and
\begin{equation}
Y_{i\sigma}(\vec{x})= \int^\Lambda \frac{d\vec{p}}{(2\pi a )^2} e^{
- i \vec{p}\cdot \vec{x}} \; Y_{i\sigma} ((-1)^{i+1} \vec{K}+\vec{p}).
\end{equation}
$Y=u,v$ are the electron annihilation operators on sublattices $A$ and $B$, respectively. In this representation, the relativistically invariant Hamiltonian takes the form $H_D=I_2 \otimes i \gamma_0 \gamma_i \hat{p}_i$. The four-component mutually anticommuting Hermitian $\gamma$-matrices belong to the ``graphene representation", $\gamma_0 = I_2 \otimes \sigma_3$, $\gamma_1= \sigma_3 \otimes \sigma_2$, $\gamma_2= I_2 \otimes \sigma_1$, $\gamma_3= \sigma_1 \otimes \sigma_2$, and $\gamma_5= \sigma_2 \otimes \sigma_2$. Here, $\left(I_2,\vec{\sigma} \right)$ are the standard Pauli matrices. The chiral $U_c(4)$ symmetry of $H_D$ is generated by $\left(I_2,\vec{\sigma} \right) \otimes \left(I_4,\gamma_3,\gamma_5,i \gamma_3 \gamma_{5} \right)$ \cite{herbut-juricic-roy}. Three generators ofrotation of electron spin are $\vec{S}=\vec{\sigma} \otimes I_4$. 
\\

The free Dirac Hamiltonian in the presence of both real ($A_i$) and pseudo ($a_i$) gauge potentials reads \cite{consult}
\begin{equation}
H_D\left[ A, a \right]=I_2 \otimes i\gamma_0 \gamma_i \left( \hat{p}_i -A_i -i \gamma_{3} \gamma_5 a_i \right).
\label{DiracHamilmagneticfields}
\end{equation}                
The real (pseudo) magnetic field breaks (preserves) the time reversal symmetry (TRS), represented by $I_t=\sigma_2 K \otimes i\gamma_1 \gamma_5 K$ in ``graphene representation" and $K$ is the complex conjugate \cite{herbut-juricic-roy, godfried}, but preserves (breaks) \emph{chiral} symmetry (CS). Separately, both real and pseudo magnetic fields quench the linear spectrum of Dirac quasi-particles into a set of Landau levels at well separated energies $\pm \sqrt{2 n B (b)}$, with $n=0,1,2, \cdots$, and degeneracies $ \Omega 2 \pi B (b)$, where $\Omega$ is the area of the sample. However, the states in zeroth Landau level near two Dirac points reside on the complimentary and same sub-lattices, respectively. The distinct natures of the zeroth Landau level can lead to different broken symmetry phases near and at the charge-neutrality point \cite{aswin, roy-thesis}. On the other hand, in the presence of a real as well as a pseudo magnetic field, the spectrum of the Dirac quasi-particle is composed of two \emph{inequivalent} interpenetrating sets of Landau levels at energies $\pm \sqrt{2 n |B \pm b|}$, with degeneracies $\Omega \pi |B \pm b| $, respectively, residing in the vicinity of two Dirac points \cite{free-LL}. Next, we consider vortex defects of various underlying U(1) symmetric order-parameters in graphene, subject to real and pseudo magnetic fields and study the zero energy modes bound to them.  
\\
 
\section{Vortex with easy-plane Anti-ferromagnet order}

The onsite Hubbard repulsion is the strongest interaction in graphene \cite{katsnelson} and favors AF order at neutral filling if sufficiently strong \cite{catalysis}. Hence, we consider a uniform background of electron density and staggered magnetization. Therefore, after keeping the only relevant term, the onsite Hubbard interaction reads as 
\begin{equation}
H_U=- \frac{U}{16} \: [ \vec{f}(\vec{A} ) - \vec{f} (\vec{A}+\vec{b})]^2. 
\end{equation}
Here, $\vec{f} (\vec{X})  = c^\dagger_\sigma (\vec{X}) \vec{\sigma}_{\sigma \sigma'}  c _{\sigma'} (\vec{X}) $ is the average magnetization on sublattice $X= A, B$, defined in terms of $c_\sigma=u_\sigma(\vec{A}), v_\sigma(\vec{B})$, respectively. The effective single-particle Hamiltonian in a fixed N\'{e}el background is 
\begin{equation}
H_N\left[ A,a \right]= H_D\left[ A,a \right]-\left( \vec{N} \left( \vec{x} \right)\cdot \vec{\sigma} \right) \otimes \gamma_0,
\end{equation}
where $ \vec{N} = \langle \vec{f}( \vec{A} )- \vec{f} (\vec{A}+\vec{b})\rangle \neq 0$ \cite{herbut-AFvortex, KKprimemixing}. In the absence of gauge fields, a constant N\'{e}el order, $\vec{N}(\vec{x})=\vec{N}$, leads to a gapped spectrum for the relativistic quasiparticles, $E=\pm \sqrt{k^2+|N|^2}$, and the order parameter ($\vec{N}$) breaks the CS \cite{herbut-juricic-roy}. In the presence of real magnetic fields, a finite N\'{e}el order shifts the Landau levels at finite energies to $\pm \sqrt{2 n B + N^2}$ with degeneracies $2 \pi B$, whereas, the zeroth Landau level gets splits to $\pm |N|$ with degeneracy $\pi B$, per unit area \cite{herbut-QHE}. Next, we assume $N_3(\vec{x})=0$ and $N_1 (\vec{x})+i N_2 (\vec{x})=|\Delta(r)| e^{i \; p \;\phi}$, $p=\pm 1$: vortex configuration \cite{zeemaneasyplane}. Here, we use a rotationally invariant representation of spinor $\Psi_\sigma(\vec{x})$. Thus the choice of the easy plane is arbitrary and does not affect the outcomes. 
\\

In the absence of gauge fields, there exist two zero energy modes in the spectrum of $H_N[0,0]$ \cite{herbut-AFvortex}. Neither $H_D[A,a]$ nor $\gamma_0$ mixes two inequivalent Dirac points, hence, $H_N [A,a]$ is block diagonal in the \emph{valley index}. One can, therefore, cast $H_N[A,a]$ as $H_1 \oplus H_2$, by exchanging the second and third $2 \times 2$ blocks. Both $H_1$ and $H_2$ however, are unitarily equivalent to a generic Dirac Hamiltonian with \emph{mass-vortex} in the presence of different \emph{effective} magnetic fields, 
\begin{equation}
H=H_0(A^t)+N_1(\vec{x}) i\gamma_0 \gamma_3 + N_2(\vec{x}) i\gamma_0 \gamma_5,
\label{massvortexfields}
\end{equation} 
where $H_0(A^t)=i\gamma_0 \gamma_i (-i\partial_i-A^t_i)$. Specifically, $H_1=U_1^\dagger H U_1$, with $U_1=I_2\oplus i\sigma_2$ and $A^t_i=A_i+a_i$, whereas $H_2=U_2^\dagger H U_2$, with $U_2= i\sigma_2 \oplus I_2$ and $A^t_i=A_i-a_i$ \cite{bitan-oddQHE}. Therefore, the effective single-particle Hamiltonian with a vortex for the easy plane of the N\'{e}el vector is equivalent to \emph{two} copies of the Dirac Hamiltonian with a twisted mass (vortex), introduced by Jackiw and Rossi \cite{Jackiw-topo} and recently studied in the context of Kekul\'{e} orders in graphene \cite{chamon-hou-mudry}, but subject to effective magnetic fields \cite{herbutmassmagnet}. These two copies, however, experience different effective fields. For example, choosing $\chi_{A}(r)=B r/2$ and $\chi_{a}(r)=b \; r/2$, so that $\vec{A}(\vec{r})=\chi_{A}(r)(\hat{y},- \hat{x})$ and $\vec{a}(\vec{r})=\chi_{a}(r)(\hat{y}, -\hat{x})$ yields uniform magnetic fields $B \pm b$ for $H_{1(2)}$. Since the pseudo magnetic fields can be produced by deliberately buckling the graphene flake, it is likely to be \emph{non-uniform}, in general. Nevertheless, a previous study always guarantees the existence of a zero energy modes in the spectrum of $H_1$ and $H_2$, irrespective of the modulation of the fields and/or vortex profile \cite{herbutmassmagnet}. Therefore, $H_N[A,a] \equiv H_1 \oplus H_2$ always hosts two zero modes. Encouraged by a recent experiment \cite{Levy}, we can take the axial field to be uniform. The characteristic lengths of two zero modes depend on the effective \emph{uniform} magnetic fields. Therefore the zero energy sub-space is composed of two modes with different length scales. Let us consider an extreme limit: $B \pm b \gg \Delta(r)= \Delta_0$ (constant). To the leading order, one can write the localization lengths of the zero modes as
\begin{equation}
L_\pm = \frac{1}{\sqrt{B \pm b}} + {\cal O}(\frac{\Delta^2_0}{B \pm b}).
\label{characlength}
\end{equation}
This result physically makes sense, since with a weak order parameter, to the leading order, magnetic fields determine the characteristic length scales. Next, we consider a prototype vortex profile, $|\Delta(r)|=r \Delta_0/R$ when $r<R$ and $\Delta_0$ (constant) when $r>R$ and the fields to be uniform \cite{herbutmassmagnet} and compute the zero energy modes explicitly and thereby $L_{\pm}$. 
\\

\section{zero modes in uniform fields}

Before we arrive at the zero modes for the easy-plane N\'{e}el vortex, it is worth studying the zero modes with the Kekul\'{e} mass vortex in the presence of an effective \emph{real} magnetic field. Even though this problem has been studied previously \cite{herbutmassmagnet}, it is worth reviewing the solution for the sake of completeness of the discussion. The equations for the zero energy mode, $\Psi_0(\vec{x})=(u_+,v_+,u_-,v_-)^\top$, of the massive Dirac Hamiltonian, 
\begin{equation}
H= i \gamma_0 \gamma_i \left( \hat{p}_i -A^t_i \right) + |\Delta(r)| \left( i \gamma_0 \gamma_3 \cos{\phi} + i \gamma_0 \gamma_5 \sin{\phi} \right)
\label{vortexmagnet}
\end{equation}
with a mass vortex and effective magnetic field reads as
\begin{eqnarray}
\partial_r  q(r) &=& - |\Delta(r)| p (r) - \chi^t(r) q(r), \\
\partial_r  p(r) &=& - |\Delta(r)| q (r) + \chi^t(r) p(r).
\end{eqnarray} 
Here we choose a symmetric gauge $\vec{A}^t(\vec{r})=\chi^t(r)(\hat{y},- \hat{x})$ and redefine the spinor components as $u_+(\vec{r})=\sqrt{i} \;p(r)$ and $u_-(\vec{r})=\sqrt{-i}\; q(\vec{r})$, with $v_{\pm}=0$. The other possibility with $u_{\pm}=0$, does not lead to a normalizable zero energy state as in the absence of fields \cite{chamon-hou-mudry, herbut-AFvortex}. For example, assuming a finite field at the origin, i.e., $\chi(r) \propto r$, yields $v_\pm \propto 1/r$ near the origin. Therefore, $v_\pm$ cannot be normalized. To find the zero modes, let us define a time like variable, $dt=|\Delta(r)|\; dr$, as in Ref. \cite{herbutmassmagnet}. One can then identify the above two equations as the Hamilton's equations, previously argued in Ref. \cite{herbutmassmagnet}, for the canonical momentum $(p(r))$ and position $(q(r))$. The corresponding time varying classical Lagrangian,
\begin{equation}
-{\cal L}(q,\dot{q},t)=\frac{1}{2}\dot{q}^2+ \frac{\zeta^t(t)}{2} q^2+\frac{1}{2} \;\frac{d}{dt} \left[ f^t(t)\; q^2 \right],
\end{equation}  
describes the motion of a classical particle in a repulsive harmonic potential (upto a total time derivative), $\zeta^t(t)=1+f^t(t)^2-\dot{f}^t(t)$, where $f^t(t)=\chi^t(t)/|\Delta(t)|$. In the presence of magnetic fields the potential gets steeper with time, otherwise, it remains static. In the mechanical analogy, one may understand the zero energy mode as the trajectory of a particle starting far from the origin with the right amount of energy, so that it reaches the origin with zero velocity at infinite time. Since the particle with sufficiently large (small) initial energy flies (goes back) to negative (positive) infinity, continuity of solutions, therefore guarantees the existence of the zero energy mode in the spectrum $H$ in Eq. [\ref{vortexmagnet}]. \\

The vortex Hamiltonian with the easy plane components of the N\'{e}el order, subject to magnetic fields, is shown to be unitarily equivalent to two copies of $H$ in Eq. [\ref{vortexmagnet}], namely, $H_1 \oplus H_2$. However, $\chi^t(r)=\chi_A(r)+\chi_a(r)$ for $H_1$ and  $\chi^t(r)=\chi_A(r)-\chi_a(r)$ for $H_2$. Here, $\chi_A(r)$ and $\chi_a(r)$ are the vector potentials of the real and the pseudo magnetic fields, respectively. The above mentioned mechanical analogy of the zero modes ensures the existence of \emph{two} zero energy states, irrespective of the field and the vortex profile \cite{herbutmassmagnet}.
\\

Next, we consider $\chi^t(r)=(B \pm b)r/2$ for $H_{1(2)}$, respectively. Therefore, the effective fields read as $B_{eff}= \left( \partial_r + 1/r \right) \chi^t(r)$, yielding $B_{eff}=B+b\equiv B_+$ for $H_1$ and $B-b \equiv B_-$ for $H_2$. Let us take the vortex configuration as in Ref. \cite{herbutmassmagnet}. $|\Delta(r)| \:=\:\Delta_0 r/R$ for $r<R$ (vortex-core) and $\Delta_0$ for $r>R$. For $r\leq R$, the components of two zero modes, say $\Psi_{1,0}$ and $\Psi_{2,0}$, are 
\begin{equation}
q^{<}_{\pm}(r)= C^{\pm}_1 e^{r^2/L^2_{\pm}} \;+\; C^{\pm}_2 e^{-r^2/L^2_{\pm}},
\label{q-origin}
\end{equation}         
and canonical momentum $p(r)=\partial {\cal L}(q,\dot{q},t)/\partial \dot{q}$ is  
\begin{equation}
p^{<}_{\pm}(r) =\frac{-2 R r}{\Delta_0 L^2_\pm}\left[ C^{\pm}_1 e^{r^2/L^2_{\pm}} \;-\; C^{\pm}_2 e^{-r^2/L^2_{\pm}} \right].
\label{p-origin}
\end{equation}         
Here, $\pm$ corresponds to the zero eigenstates of $H_{1(2)}$, respectively. $C^{\pm}_i$ are constants, and $L_{\pm}$ are the characteristic lengths of the zero modes,
\begin{equation}
L^{-4}_{\pm}=\frac{1}{4} \left(\frac{\Delta^2_0}{R^2} + \frac{(B \pm b)^2}{4} \right).
\label{lengths}
\end{equation}  
Whereas, far from the core of the vortex ($r>R$)
\begin{equation}
q^{>}_{\pm}(r)=C_3 e^{-\frac{(B \pm b) r^2}{4}} F\left( \frac{3|\delta|+1-\delta}{4 |\delta|};\frac{3}{2};\frac{(B \pm b) r^2}{2}\right),
\label{deltalessone}
\end{equation}
when $\delta=B/2\Delta^2_0 < 1$. On the other hand, for $\delta > 1$
\begin{equation}
q^{>}_{\pm}(r)=C_3 e^{-\frac{(B \pm b) r^2}{4}} F\left( \frac{2\delta+1}{4 \delta};\frac{3}{2};\frac{(B \pm b) r^2}{2}\right),
\label{deltagreaterone}
\end{equation}
where $F$ is the hypergeometric function \cite{herbutmassmagnet, tableandintegral}. For $r>R$, $dt=\Delta_0 dr$, yielding 
\begin{equation}
p^{>}_{\pm}(r)\;=\; - \; (1/\Delta_0) \; (\partial q^{>}_{\pm}(r)/\partial r).
\end{equation} 
Continuity implies that the value  and the first derivative of the solutions in two regions must match at $r=R$, which eliminate two out of the three constants, whereas, the remaining one is fixed by the normalization condition. Equation (\ref{lengths}) reflects that two zero energy modes have different characteristic lengths. Otherwise two zero modes of $H_N[A,a]$ are
\begin{eqnarray}
\Psi_{1,0}&=& U^\dagger_1 \left( p_+ ,0,q_+,0\right)^\top (B_+), \\
\Psi_{2,0}&=& U^\dagger_2 \left( p_-,0,q_-,0\right)^\top (B_-),
\end{eqnarray}
where $U_1=I_2 \oplus i \sigma_2$ and $U_2= i \sigma_2 \oplus I_2$. Note $\Psi_{1,0}$ and $\Psi_{2,0}$ are the functions of the effective fields $B+b$ and $B-b$, respectively. In the following discussion, we show how the existence of two length scales yields various peculiarities. However, the following discussion is insensitive to the exact form of $L_\pm$.
\\

\section{internal structure of the zero modes }
In the presence of real and pseudo magnetic fields, the zero energy eigenstates of $H_1$ and $H_2$ are
\begin{equation}
 \Psi_{1,0}= \left( u_{1 \uparrow}, v_{1 \uparrow},  u_{1 \downarrow},v_{1 \downarrow} \right)^\top
=\sqrt{i} \left(p_+ (r),0,0,-i q_+(r) \right)^\top,
\end{equation}
and 
\begin{equation} 
\Psi_{2,0}= \left( u_{2 \uparrow}, v_{2 \uparrow},  u_{2 \downarrow},v_{2 \downarrow} \right)^\top
=\sqrt{i} \left(0,p_- (r),-i q_-(r),0 \right)^\top,
\end{equation}
respectively. $q_\pm (r)$ and $p_\pm (r)$ have been computed exactly in uniform fields with the above mentioned profile of the vortex mass. Typically, the characteristic length for $\Psi_{1,0}$ is smaller than that for $\Psi_{2,0}$, for example $L_{\pm}$ in Eq.\ (\ref{lengths}). Consequently, the former is squeezed into the vortex-core. 
\\

\begin{figure}[htb]
\includegraphics[width=6.cm, height=2.75cm]{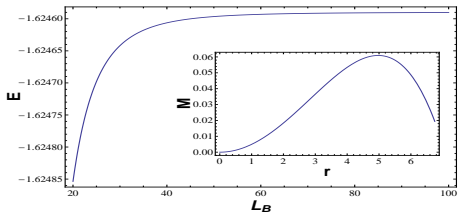}
\caption{Energy $(E/N_3)$ as a function of magnetic length ($L_B/a$), with $B(r)=B$ and $N_3(r)=N_3$ for $r\leq R$, zero elsewhere. The vortex configuration is $\Delta(r)=\Delta_0$ for $r\geq R$, otherwise zero. We set $R=5 \mathring{A}$ and $\Delta_0=30$ eV. Inset: Absolute magnetization ($|M/m_3|$) as a function distance (in $\mathring{A}$) from the center, for $L_B=40 \mathring{A} \sim 1/\sqrt{B}$.}\label{energy}
\end{figure}

The magnetization within the zero energy sub-space is
\begin{equation}
H_Z= \lambda \: \left( \sigma_3 \otimes I_2 \right) \oplus \left( \sigma_3 \otimes I_2 \right),
\end{equation}
proportional to the Zeeman term. $\lambda = g \: B(\vec{x})$ and $g \approx 2$ for electrons in graphene. In the absence of fields, $p_\pm(r)=q_\pm(r)=\exp[-\int^r_0 |\Delta(z)| \; dz]$. Each of the states then possesses zero magnetization everywhere in the space \cite{herbut-AFvortex}. In the presence of fields, the radial dependences of $q(r)$ and $p(r)$ are different, in general. Therefore, the zero modes can exhibit finite but \emph{oscillatory} magnetization, shown in the inset of Fig. \ref{energy}. The overall net magnetization is still zero for each state and zero energy sub-space remains unperturbed. Such a \emph{local} magnetic moment can be probed by \emph{magnetic force microscope} (MFM) measurements. Replacing the vortex by an anti-vortex simply exchanges the role of $u$ and $v$ components. For $b>B$, $\Psi_{1,0}$ remains invariant, but the role of $q_-(r)$ and $p_-(r)$ is exchanged in $\Psi_{2,0}$. This limit seems quite conceivable, since, in the experiment, $b \sim 350$ T \cite{Levy}, but the highest laboratory magnetic field is $\sim 45$ T. \\

Two zero energy modes $\Psi_1=(\Psi_{1,0},0)$ and $\Psi_2=(0,\Psi_{2,0})$ comprise a two dimensional basis in the zero energy manifold (${\cal H}_0$). Any operator that commutes or anticommutes with $H_N[A,a]$ leaves that space invariant. When $a=0$, there are four matrices falling in the second category, namely $\left(\sigma_3\otimes\gamma_0, I_2 \otimes i \gamma_0 \gamma_3, I_2 \otimes i \gamma_0 \gamma_5, \sigma_3 \otimes i \gamma_1 \gamma_2\right)$, which together, close a $Cl(3)\times U(1)$ algebra \cite{herbut-topo, group}. $\sigma_3 \otimes \gamma_0$ is the z-component of the N\'{e}el order and the remaining members of the $Cl(3)$ group correspond to the KBDW. The U(1) part is generated by $\sigma_3 \otimes i \gamma_1 \gamma_2$, representing the third component of the spin-triplet version of the time-reversal symmetry-breaking order \cite{raghu}. When the pseudo magnetic field is finite, the KBDW orders no longer anticommute with $H_D[A,a]$. Nevertheless, the expectations value of all \emph{four} orders come only from ${\cal H}_0$. The expectation value of any physical observable is
\begin{equation}
\langle m(\vec{x}) \rangle=\frac{1}{2}\left( \sum_{occup}-\sum_{empty} \right) \Psi^\dagger_E (\vec{x})  M \Psi_E (\vec{x}),
\end{equation}  
where $M$ is the traceless matrix and $\{ \Psi_E (\vec{x}) \}$ is the set of eigenstates, with eigenvalues $E$ of a generic Hamiltonian. The existence of a unitary matrix $T$, which commutes with $M$, while anticommuting with the Hamiltonian, restricts the above sum within ${\cal H}_0$ \cite{herbut-AFvortex, LCRwijewardhana}. Choosing $T=\sigma_3 \otimes \gamma_0$, which anticommutes with $H_N[A,a]$, one finds that the expectation value of either $M=\sigma_3 \otimes \gamma_0$ or $\sigma_3 \otimes i \gamma_1 \gamma_2$ is solely determined by $\Psi_{i,0}$'s. For $M=I_2 \times i \gamma_0 \gamma_5 \;\mbox{or}\; I_2 \times i \gamma_0 \gamma_5$ or any linear combination, one can choose $T=\sigma_3 \otimes i \gamma_1 \gamma_2$. Therefore, the AF and QSH as well as the KBDW orders acquire their expectation values only from the zero energy sub-space. 
\\

When the N\'{e}el vector is tilted out of the easy plane $(N_3 \neq 0)$, the zero energy manifold gets split, since $N_3 \; \langle \Psi^\dagger_{i,0} \left( \sigma_3 \otimes \gamma_0 \right) \Psi_{j,0}\rangle=\pm N_3 \delta_{i j}$ \cite{herbut-AFvortex}. Therefore, at half-filling, a finite $N_3$ keeps only one of the zero energy states occupied while leaving the other one empty. On the other hand, if the N\'{e}el order forms a half-skyrmion, $\Psi_{i,0}$'s are no longer degenerate. If $|N_3 (\vec{r})| \ll \Delta_0$, one can neglect the mixing of zero modes with the rest of the spectrum. The energy is then maximally lowered by filling the state with smaller characteristic length ($L_+$). The physical reason: $N_3(\vec{x})$ is finite in the core of the vortex and smoothly vanishes toward the boundary. Out of the two states, the one with the smaller length scale experiences a larger overlap with the out-of-plane component of the N\'{e}el order and, therefore, being filled lowers the energy maximally. The effective magnetic fields for $\Psi_{1,0}$ and $\Psi_{2,0}$ are $B+b$ and $|B-b|$, respectively. The characteristic length is smaller for $\Psi_{1,0}$. Hence, state $\Psi_{1,0}$, is occupied in the presence of a weak half-skyrmion or meron configuration of $\vec{N}$. This feature is independent of specific configuration of field or vortex or $N_3$. Nevertheless, energy as a function of magnetic length, for a chosen vortex profile, $N_3$ and the fields is shown in Fig. \ref{energy}.
\\

\section{Easy-plane quantum spin Hall vortex}
If, on the other hand, repulsion among the fermions living on the second-neighbor sites of the honeycomb lattice is sufficiently strong, a QSH order $(\vec{C}=\langle \sum_\sigma \Psi^\dagger_\sigma [ \vec{\sigma} \otimes i \gamma_1 \gamma_2] \Psi_\sigma \rangle \neq 0)$ can be realized in graphene. It breaks the TRS for each spin component and corresponds to circulating currents among the sites of the same sub-lattice \cite{raghu}. However, the QSH phase can be realized even at infinitesimal next-nearest-neighbor interactions in the presence of finite pseudo magnetic fields \cite{herbut-pseudo, bitan-oddQHE, aswin, roy-thesis}. The single-particle Hamiltonian with an underlying QSH order reads as
\begin{equation}
H_C\left[ A,a \right]= H_D\left[ A,a \right]-\left( \vec{C} \left( \vec{x} \right)\cdot \vec{\sigma} \right) \otimes i \gamma_1 \; \gamma_2.
\end{equation}
Alike AF, the QSH order is also block diagonal in the valley index and can be projected onto the easy plane by a finite Zeeman coupling \cite{zeemaneasyplane}. Otherwise, $N_i \rightarrow (-1)^{j+1} C_i$ for $H_j$ with $i, j=1,2$ \cite{KKprimemixing}. The two zero energy modes are the following: $\Psi_{1,0}$ remains unchanged, while $p_-(r) \leftrightarrow q_-(r)$ in $\Psi_{2,0}$. Within ${\cal H}_0$, the members of the $Cl(3)$ algebra are $\left( \sigma_3 \otimes \gamma_0, \sigma_3 \otimes i \gamma_0 \gamma_3, \sigma_3 \otimes i \gamma_0 \gamma_5 \right)$. The last two entries corresponds to the z-component of the spin-triplet KBDW. The U(1) component is formed by $\sigma_3 \otimes i \gamma_1 \gamma_2$. A finite third component of the QSH order ($C_3$) can be developed in the core by making the vortex charged, since it acts like an \emph{identity} operator in ${\cal H}_0$. An out-of-plane component of the N\'{e}el order splits ${\cal H}_0$. Therefore, onsite Hubbard repulsion (U) can lift the degeneracy of the zero energy sub-space by developing a finite $N_3$ in the core of the vortex.
\\

It is worth confirming that all four mass-orders, constituting the $Cl(3) \times U(1)$ algebra, acquire their expectation values from ${\cal H}_0$, even though some of the members do not anti-commute with the Hamiltonian upon imposing an axial magnetic field. If we select $M=\sigma_3 \otimes \gamma_0$ or $\sigma_3 \otimes i \gamma_1 \gamma_2$, one can choose $T= \sigma_3 \otimes \gamma_0$. And for $M= \sigma_3 \otimes i \gamma_0 \gamma_3 $ or $\sigma_3 \otimes i \gamma_0 \gamma_5$, one can again choose $T=\sigma_3 \otimes i \gamma_1 \gamma_2$, same as before.            
\\

\section{Kekul\'{e} vortex}
In contrast to the AF and QSH orders, the KBDW couples two inequivalent Dirac points and breaks the translational symmetry of the lattice into the Kekul\'{e} pattern \cite{chamon-hou-mudry, herbut-roy-half-vortex}. In the presence of gauge fields, the effective single-particle Hamiltonian $H_{K}[A,a]$ with an underlying vortex of KBDW order is 
\begin{equation} 
H_D[A,a]-m_r I_2 \otimes  i \gamma_0 (\gamma_3 \cos{\phi} + \gamma_5 \sin{\phi} ) \equiv U H_{K}[A,0] U,
\end{equation}
where $U=\exp[{-\rho(\vec{x}) ( I_2 \otimes \gamma_0) }]$. $m_r$ counts the radial variation of the vortex profile. $I_2 \otimes \gamma_0$ anticommutes with $H_{K}[A,0]$ and acts like an identity matrix in ${\cal H}_0$. The zero energy states in the presence of the pseudo field, therefore, differ from the one in its absence by a factor exp$[\rho(\vec{x})]$, where $b=\epsilon_{ij} \partial_i a_j=\partial^2 \rho(\vec{x})$ \cite{jackiw-Pi}. However, two zero modes are identical for each spin component and have \emph{equal} characteristic lengths. The explicit form of the zero modes without the pseudo fields can be computed simply by setting $b=0$ in the previous exercises. 
\\

The exclusive members of the $Cl(3)$ algebra of the masses are $\vec{N}=\left( \vec{\sigma} \otimes \gamma_0 \right)$, three components of the N\'{e}el vector, whereas, that of the U(1) part is $I_2 \otimes \gamma_0:$ charge-density-wave order. The Zeeman coupling $\lambda \left( \sigma_3 \otimes I_4 \right)$ is proportional to $N_3$ within ${\cal H}_0$, thus may develop its finite expectation value by splitting ${\cal H}_0$. On the other hand, a charge-density wave order can be developed inside the core of the vortex if it is charged. Contrary to the previous examples, all \emph{four} orders anti-commute with the Hamiltonian $H_K[A,a]$ even when an axial field is present. One can choose $T= I_2 \otimes \gamma_0$, for all four order parameters. Hence, their expectation values can be computed only from the zero energy states.   
\\

\section{Summary and discussions}
To summarize we find that in the presence of real and pseudo magnetic fields, all the U(1) symmetric orders in graphene can host two zero modes with underlying vortex defects. However, two zero modes can have either one or two characteristic lengths, depending on the nature of the underlying insulating orders. We show that, if the underlying order parameter couples two inequivalent Dirac points, two zero modes have the same characteristic length, whereas, with translational symmetric order parameters, zero modes enjoy different length scales. The existence of two zero modes leads to additional competing order parameters in the core of the vortex. These order parameters together close a $Cl(3) \times U(1)$ algebra, and their expectation values \emph{always} arise only from zero energy sub-space. When the mid-gap states have different length scales, e.g., with underlying AF and QSH orders, charge is continuous and \emph{oscillatory}; the overall neutrality of the system is preserved only far from the vortex core. Mid-gap states with identical length scales do not lead to any excess charge anywhere in the space. The excess \emph{local} charge can be measured by a \emph{scanning tunnel microscope} (STM) probe. It can, therefore, also serve the purpose of litmus test to determine the possible broken translational symmetry in the system. Zero modes with different scales can be delocalized from each other by tuning the relative strength of two magnetic fields, thereby keeping a \emph{single} state in the core. For example, when $B \approx b, \Psi_{1,0}$ is highly localized near the vortex core. However, $\Psi_{2,0} \approx \exp{(-\int^{r}_0 |\Delta(t)| dt)}$ experiences effectively zero field, therefore, exists even beyond the vortex core. Moreover, if $B,b \gg \Delta_0/R,$ $q_+(r)$ is localized very close to the vortex core, and $p_+(r) \sim (B+b) R/\Delta_0 q_+(r) \gg q_+(r)$ \cite{herbutmassmagnet}. A finite magnetization, therefore, persists beyond the vortex core and points opposite to that in the core. Net magnetization is still zero.
\\ 

Besides the insulating orders, one can also consider the underlying vortex of various superconducting orders (SCOs) in graphene. Aside from the Kekul\'{e} superconductors, the rests, e.g., s-wave or f-wave SCOs, couple two inequivalent Dirac points and may not host \emph{Majorana} modes \cite{ghaemi, majorana}. However, a Kekul\'{e} SCO with topological defects, e.g., vortex or half-vortex can lead to Majorana modes \cite{herbut-roy-half-vortex}. Those Majorana modes are valley polarized. Hence, in the presence of real and pseudo magnetic fields, the Majorana modes experience different effective magnetic fields. Therefore, by adjusting the relative strength of the fields, one can localize one of the Majorana modes within the vortex core, while the other one can be pushed outside. Binding of a single Majorana mode can also be useful for quantum computations. Otherwise, the existence of zero modes in the presence of vortices can lead to KT scaling of the longitudinal resistivity $(R_{xx})$ in a neutral graphene ($\nu=0$), subject to real \emph{and} pseudo magnetic fields \cite{bitan-oddQHE}. Zero modes with different length scales can also be found in the spectrum of a \emph{birefringent} Dirac fermion with a kinked staggered chemical potential \cite{malcolm}. In a recent paper \cite{roy-smith-kennett}, authors considered dual vortices: a vortex in the birefringent parameter as well as the standard mass vortex. Under that circumstance, vortex and anti-vortex zero modes decay with different length scales.
\\

\section{Acknowledgements}
This work was supported by the NSERC of Canada. The author is grateful to Igor Herbut, Erol Girt, Malcolm Kennett, Oskar Vafek, and Nick Bonesteel for useful discussion. The author is thankful to Payam Mousavi and Vladimir Cvetkovic for  critical reading of the paper.


\begin{thebibliography}{99}
\bibitem{consult} A general $SU(2)$ pseudo gauge potential is $a_i=\gamma_3 a^3_i + \gamma_5 a^5_i + i \gamma_3 \gamma_5 a^{35}_i$ \cite{bitan-oddQHE}. However, smoothly deformed graphene flake does not break the translational symmetry \cite{Levy}, generated by $I_2 \otimes i \gamma_3 \gamma_5$ \cite{herbut-juricic-roy}. The first two entries couples two Dirac points and breaks the translational symmetry. Thus we set $a^3_i=a^5_i=0$ and $a^{35}_i \equiv a_i$.
\bibitem{catalysis} In the presence of magnetic fields ordered phases can be realized even at infinitesimal interactions. See \cite{bitan-oddQHE, herbut-juricic-roy, herbut-pseudo}.
\bibitem{KKprimemixing} In general, Hubbard interaction contains terms with both inter and intra valley scattering. In the presence of magnetic field however one is at luxury of keeping the least irrelevant term after integrating out the Fourier modes between $1/a < |\vec{k}| < 1/{L_B}$, which is the Gross-Neveu (GN) term $\left[ \sum_\sigma \Psi^\dagger_\sigma (\vec{\sigma} \otimes \gamma_0 ) \Psi_\sigma \right]^2$. Similarly while considering the next neighbor repulsion, we keep the other GN term $\left[ \sum_\sigma \Psi^\dagger_\sigma (\vec{\sigma} \otimes i \gamma_1 \gamma_2 ) \Psi_\sigma \right]^2$. This terms do not couple two inequivalent Dirac points. $L_B :$ magnetic length.
\bibitem{zeemaneasyplane} A finite Zeeman coupling, for example $\lambda \: ( \sigma_3 \otimes I_4 )$ commutes with the Dirac Hamiltonian and the easy axis AF/QSH orders, while anti-commuting with their easy plane components. Energy of the filled Dirac-Fermi sea is therefore maximally lowered by projecting the N$\acute{e}$el/QSH order into an easy plane.
\end{thebibliography}
\end{document}